\documentclass[aps,prl,twocolumn,showpacs]{revtex4-1}
\usepackage{amsmath, amssymb}
\usepackage{amsfonts}
\usepackage{graphicx}
\usepackage{bm}
\usepackage{color}
\renewcommand{\d}[2]{\frac{\mathrm{d} #1}{\mathrm{d} #2}} 

\begin{document}

\title{Nonlinear focusing in dynamic crack fronts and the micro-branching transition}
\author{Itamar Kolvin$^1$, Jay Fineberg$^1$ and Mokhtar Adda-Bedia$^2$}

\affiliation{$^1$ Racah Institute of Physics, The Hebrew University of Jerusalem, Jerusalem, Israel 9190401 \\
$^2$  Universit\'{e} Lyon, Ecole Normale Sup\'{e}rieure de Lyon, Universit\'{e} Claude Bernard, CNRS, Laboratoire de Physique, F-69342 Lyon, France }

\date{\today}

\begin{abstract}
Cracks in brittle materials produce two types of generic surface structures: facets at low velocities and micro-branches at higher ones. Here we observe a transition from faceting to micro-branching in polyacrylamide gels that is characterized by nonlinear dynamic localization of crack fronts. To better understand this process we derive a first-principles nonlinear equation of motion for crack fronts in the context of scalar elasticity. Its solution shows that nonlinear focusing coupled to rate-dependence of dissipation governs the transition to micro-branching.
\end{abstract}

\pacs{}
\maketitle

Fracture is typically an irregular process. Cracks show a strong tendency for 
instability, creating non-smooth surfaces with rich structure. 
Crack instabilities and their associated structure exhibit a strong dependence
on crack velocity: slow tensile cracks ($v\ll c_R$, where $c_R$ is the Rayleigh wave speed) are prone to nucleate steps which drift along the crack front and divide the fracture surface into facets~\cite{Kermode.08,Kermode.13,Baumberger.08,Tanaka.98,Gent.84,Kolvin.17}; 
faster tensile cracks are unstable to the formation of micro-branches --- microscopic cracks that branch off the main crack front~\cite{Fineberg.15,Goldman.15,Livne.05,Sharon.96,Kolvin.15}. 
Linear perturbation theory, however, predicts that \textit{any} initial disturbance to a tensile crack front should either decay as the crack progresses~\cite{Gao.89,Ball.95,Ramanathanquasistatic.97} or disperse as outgoing waves \cite{Ramanathan.97,Adda-Bedia.13}. Current linear theories are therefore incapable of reproducing the observed fracture surfaces.

In recent non-perturbative approaches to fracture, such as lattice models \cite{Heizler.15,Heizler.17} and phase-field models \cite{Henry.13,Bleyer.17,Chen.17}, localized crack branching arises naturally, regardless of the specific dissipative process. Both approaches predict that micro-branching is governed by the microscopic dissipation length-scale and that instability initiates at $v_c\sim 0.7 c_R$. This critical velocity is significantly higher than that observed in experiments, and predicted from energy considerations in the theory of 2D branching \cite{Eshelby.99,Katzav.07}. In addition, micro-branch dimensions typically exceed the process zone size by a few orders of magnitude. It therefore remains unclear what component or mechanism is missing in the existing models.    

\begin{figure}[h!]
	\includegraphics[scale=1]{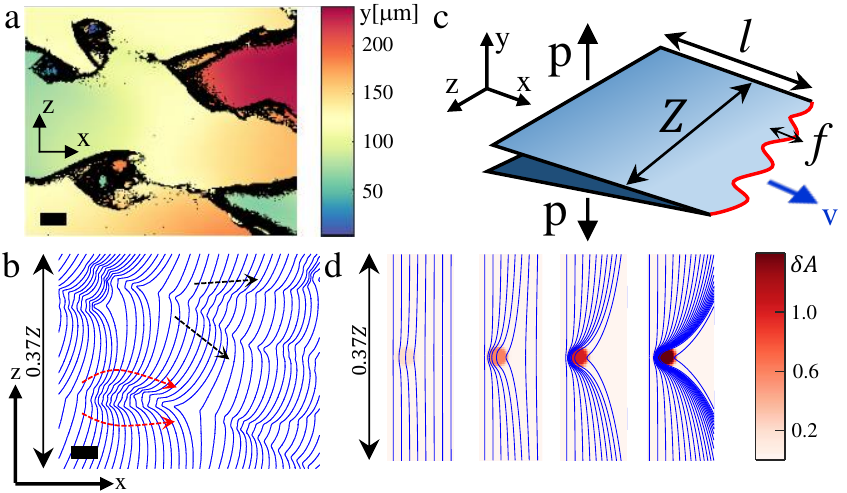}
	\caption{Experimental fracture surface (a) formed by tensile crack fronts (b) moving at $\sim 0.06 c_R$ and displayed at $0.13\, ms$ intervals. The surface features both facets (upper and lower right) and micro-branches (upper and lower left). A pair of step-lines nucleate and diverge forming a facet (black arrows) while a micro-branching event ends with cusp formation (red arrows). The black scale bars are $200\,\mu m$ long. (c) Geometry of the model. Two line loads $p$ moving with a speed $v$ are driving a planar crack front at a distance $l$. The crack front is perturbed around a straight configuration with amplitude $f(z,t)$. (d) Crack fronts, calculated via Eq.(\ref{energyreleaserate}), move at velocity $v=0.05c$ and encounter obstacles with increasing toughness; see text following Eq. (\ref{Gamma}) for details. Fronts are plotted at the same time intervals as in (b) when identifying $c=c_R$.
	}\label{fig1}
\end{figure}

In polyacrylamide gels \cite{Kolvin.17} cracks exhibit a transition between facet formation and micro-branching at $v\sim0.05-0.1c_R$.  The transition is not sharp, and both types of structures may coexist in the transition region.
A typical fracture surface (for $v \sim 0.06 c_R$) shown in Fig. \ref{fig1}(a) features both micro-branches and facets. 
 Fig. \ref{fig1}(b) depicts the corresponding crack fronts that formed these structures: 
 in the upper right part of the panel, a facet forms by a pair of steadily diverging steps. Each step induces a cusp-like deformation of the crack front. 
 In the lower left part of the panel, a micro-branch similarly initiates as a pair of diverging steps, that gradually change their direction and converge.
Although the global conditions were identical, the two structures did not meet the same fate. 

How does the crack front ``decide" which structure will eventually form? A close inspection reveals that micro-branch formation involves stronger front curvatures than facet formation.  
Hence, micro-branches embody stronger perturbations to the crack front. If the transition to micro-branching depends on the amplitude of the perturbation, it cannot be captured by a linear theory. We are, therefore, driven to consider nonlinear perturbations to dynamic crack fronts. 
 
 In this Letter we derive and solve an equation of motion for planar crack fronts that is exact to the 2\textsuperscript{nd} order in front perturbation.
In a homogeneous and isotropic medium crack front motion is dictated by specifying the local normal velocity $v_\perp$ \cite{Hodgdon.93}.  
Cracks propagate when the energy flow into the crack per unit area, $G$ or energy release rate, is balanced by the dissipation $\Gamma$ involved in creating the two new fracture surfaces.
 Energy flow into the crack tip is regulated by a universal function $g(v_\perp)$, which approaches $1$ when $v_\perp\rightarrow 0$ and $0$ when $v_\perp\rightarrow c_R$~\cite{Freund.90}. 
 Consider a straight crack front confined to the $xz$ plane and propagating along the $x$ axis with velocity $v$. When the front is perturbed, i.e. its position is given by $x=vt+f(z,t)$ (see Fig. \ref{fig1}(c)), local energy balance $G=\Gamma$ reads
\begin{equation}\label{energyreleaserate}
G_0\, g(v_\perp)\left(1+H[f]\right) = \Gamma(x,z;v_\perp)\,,
\end{equation}  
where $G_0$ is the energy release rate of the unperturbed crack. The perturbation $f$ introduces a term $H[f]$ which depends on the history of the crack front, and can be computed order-by-order in $f$, through a solution of the elastodynamic problem \cite{Willis.95,Willis.13,Ramanathan.97,Norris.07}. The fracture energy $\Gamma$ may vary locally along the front and depend on crack velocity \cite{BFM10,Boue.15}.

Computing $G$ requires an asymptotic solution of the 3D vector displacement field in the vicinity of the crack front. For a tensile (Mode I) crack, $G$ is given by the \textit{J}-integral involving the product of stress and displacement rate~\cite{Freund.90}. A calculation results in the expression $G=PK$, where $K=\sqrt{2\pi}\lim_{X\rightarrow 0^+}\sigma_{yy}(x,0,z)X^{1/2}$ and
$P=\lim_{X\rightarrow 0^-}u_{y}(x,0,z)(-X)^{-1/2}$ are the stress and displacement intensity factors respectively~\cite{Norris.07}. Here $\sigma_{ij}$ denotes the stress tensor, $u_i$ is the displacement vector and $X=x-vt-f$.   
The Mode I elastodynamic problem consists of simultaneously solving three decoupled scalar wave equations, with wave velocities being the longitudinal and the shear wave speeds respectively. Free boundary conditions on the crack faces, however, mix the wave polarizations, rendering the problem quite formidable. 
Nonlinear corrections to $G$ have not been explicitly computed yet in the general dynamic case \cite{Willis.13,Willis.14}.
 
Fortunately, $G$ can also be derived in a simpler analogous model which involves only a single scalar field $\phi$ satisfying a wave equation with wave speed $c$ \cite{Rice.94,Perrin.94}. In the quasistatic limit, the elastic fields in the bulk can be exactly written in terms of the scalar potential $\phi$~\cite{Adda-Bedia.06}, and on the $y=0$ boundary one obtains the simple forms $u_y=\phi$ and $\sigma_{yy} = \partial_y\phi$.
Moreover, both the scalar model and Mode I elastodynamics contain wave-like modes that maintain front coherence, and the universal function in the scalar model, $g(v)=\sqrt{(1-v/c)/(1+v/c)}$, closely resembles that of Mode I fracture \cite{Morrissey.98}.
In the following, we set $c=1$ for convenience.

Our derivation of $G$ uses a matched asymptotic expansion (MAE) approach \cite{Norris.07}. Here we present only a sketch of the derivation, which will be provided in detail elsewhere \cite{Kolvin.later}. 
Fig. \ref{fig1}(c) depicts a crack front driven by two line loads $p$ located at a distance $l$ from the crack front and moving at a constant velocity $v$. 
In the absence of perturbations, the loads $p$ cause the crack front to propagate steadily at a velocity $v$.
Since our problem is symmetric with respect to the $y=0$ plane, the scalar field $\phi$ can be determined by solving the wave equation in the $y>0$ half-space with zero displacement $\phi|_{y=0} = 0$ for $X>0$ and zero stress $\partial_y\phi|_{y=0} =0 $ for $X<0$.

The MAE consists of matching solutions at two scales of the problem. Assuming that $f \ll l$, we first find an ``inner" solution for $\phi$ in the vicinity of the crack front written as an expansion in powers of $X$ and of $f$. The inner solution contains integration constants that should be determined by matching to an ``outer" solution dominated by the length scale of the system $l$. In the inner solution, the most singular power must be $\phi\sim X^{1/2}$ since it ensures a finite $G$. Far from the crack front, however, the expansion $X^{1/2}\simeq \sqrt{x-vt}-\frac{1}{2}\frac{f}{\sqrt{x-vt}}-\frac{1}{8}\frac{f^2}{(x-vt)^{3/2}}+...$ only appears to create stronger ``unphysical" singularities at the location of the unperturbed front $x=vt$. The outer solution, which can be found independently of $f$, is expanded in powers of $x-vt$ and the coefficients obtained from both solutions are then compared. For each order of the perturbation only a finite number of coefficients needs to be matched. By this method, the intensity factors $P$ and $K$ are obtained from the coefficient of the $X^{1/2}$ component of the field $\phi$. 

The main result of our calculation is an expression for the history functional $H[f]$ up to second order in $f$.  First, we define the linear functionals in Fourier space $\hat{\Psi}[\hat{f}] =\alpha k\int_{-\infty}^t\frac{\mathrm{d}t'}{t-t'}\, J_1(\alpha k (t-t'))\hat{f}(k,t')$ and $\hat{\Psi}_2[\hat{f}] =\alpha^2 k^2\int_{-\infty}^t \frac{\mathrm{d}t'}{t-t'}\, J_2(\alpha k (t-t'))\hat{f}(k,t')$, where  $\alpha = \sqrt{1-v^2}$, $J_1$ ($J_2$) is the 1\textsuperscript{st} (2\textsuperscript{nd}) order Bessel functions and the hats denote the spatial Fourier transform, i.e. $\hat{f}(k,t)=\int \mathrm{d}z\, e^{-ikz}f(z,t)$. With these definitions the history functional is given in real space by
\begin{equation}\label{2ndorder}
\begin{split}
H[f] =  &-\frac{1}{\alpha^2} \Psi[f]+\frac{1}{4\alpha^4}\Psi[f]^2 +\frac{1}{2\alpha^4}\Psi[f\Psi[f]]\\
&-\frac{1-2v}{4\alpha^4}\Psi_2[f^2]-\frac{1+2v}{2\alpha^4}f\Psi_2[f]\,.
\end{split}
\end{equation}
In this expression we neglected terms of order $O(f/l)$, and consequently $H[f]$ becomes invariant to translation $f\rightarrow f+C$, where $C$ is a constant.
Taking $f$ to be time-independent and $v\rightarrow 0$, Eq. (\ref{2ndorder}) recovers, in the limit $f/l\rightarrow 0$, the corrections to $G$ calculated for quasi-static crack fronts~\cite{Vasoya.16}. 
On the other hand, assuming that $f$ does not depend on $z$, the history functional is identically zero, and Eq. (\ref{energyreleaserate}) becomes the 2D equation of motion given in~\cite{Rice.94}.

The scene is now set to test how crack fronts respond dynamically to perturbations, and determine if and how self-focusing arises due to nonlinearities.
Mimicking the experimental situation where dissipation locally increases due to the formation of surface and sub-surface structure, we write the fracture energy as a product of two parts
\begin{equation}\label{Gamma}
 \Gamma(x,z;v_\perp) = \Gamma_0(v_\perp)(1+\delta A(x,z))\,,
\end{equation}
where the ``bare" fracture energy $\Gamma_0(v)= \tilde{\Gamma}_0(1+av)$ grows linearly with crack velocity and $\delta A$ quantifies the local relative increase in fracture area.  
Expanding all quantities in Eq. (\ref{energyreleaserate}) to the 2\textsuperscript{nd} order in $f$ we obtain a nonlinear equation of motion for the crack front \cite{Supplemental}. 
The explicit equation of motion contains: a geometric term $\frac{1}{2}vf_z^2$ which accounts for propagation along the local normal to the front, elastic terms that stem from $H[f]$ and dissipative terms that depend on $\delta A$. We may now numerically propagate the crack front using an Euler scheme under periodic boundary conditions along an interval of length $Z$ (see Fig.1(c)).

We first test crack front response to a disk-shaped obstacle. $\delta A$ is taken to be constant inside the obstacle, while decaying smoothly and rapidly to zero outside. To minimize system size effects we take the diameter of the obstacle to be $d = 0.025Z$.
Fig. \ref{fig1}(d) shows crack fronts moving at $v=0.05$ and encountering obstacles with increasing toughness. The fronts were discretized over $N=512$ mesh points. As in the polyacrylamide gels used in our experiment, we assumed that the fracture energy grows linearly with velocity with $a=4$. 
As the obstacle toughness increases, the crack front dynamically develops increasingly higher curvatures while detaching from the obstacle. During detachment the elastic tension stored in the front is released, driving the crack front to accelerate to high velocities.
Increasing mesh resolution did not affect the peak curvature, indicating that elasticity arrests further curvature growth.

\begin{figure}[h!]
	\includegraphics[scale=1]{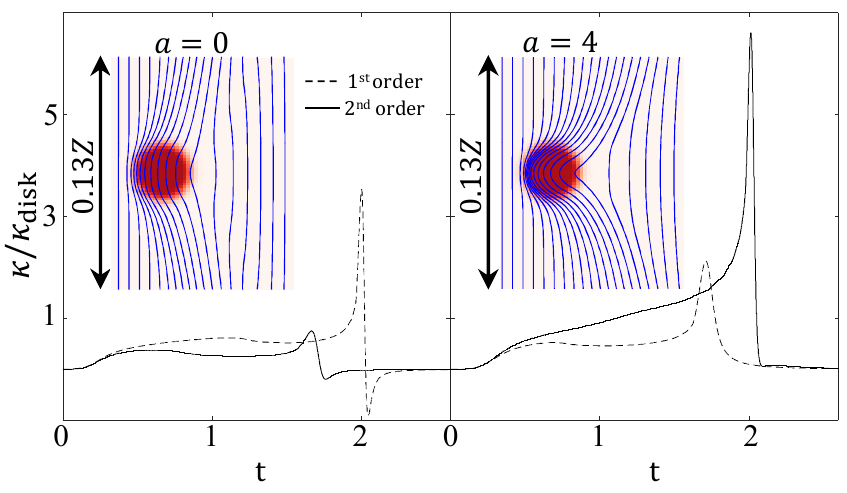}
	\caption{Change of front dynamics from defocusing to focusing when increasing $a = \d{\log\Gamma_0}{v}|_{v=0}$. Panels show curvature evolution at the obstacle mid-line for velocity-independent fracture energy (\textit{left}) and for velocity-dependent fracture energy (\textit{right}). Here $v=0.3$, $\delta A=1.2$ and $\kappa_{disk} = 2/d = 80/Z$. The 1\textsuperscript{st} order solution is drawn for comparison (dashed line). Insets depict crack front profiles drawn over the spatial fracture energy distribution.  
		}\label{fig2}
\end{figure}

Surprisingly, crack front dynamics are \textit{defocusing} when the fracture energy is velocity-independent (i.e. $a=0$). 
Fig. \ref{fig2} compares curvature evolution for $a=0$ and $a=4$ at $v=0.3$ and $\delta A=1.2$. 
When $a=4$ the front locally decelerates as it enters the obstacle and curvature builds up continuously, first gradually and then rapidly as it breaks free. 
Comparison with the solution using only 1\textsuperscript{st} order terms  shows that the nonlinearities produce curvatures up to $\sim\! 6\kappa_{disk}=12/d$.
For $a=0$, the velocity-independent case, front curvature quickly reaches a plateau at $\sim\!\kappa_{disk}/2$, while a 1\textsuperscript{st} order solution under the same conditions develops higher curvatures.   

The transition from crack front defocusing to self-focusing with increasing $a$ is a generic property of the nonlinear equation of motion. 
Eq. (\ref{energyreleaserate}) can be solved analytically in the case of a time-independent cosine perturbation $\delta A = D\cos (z)$. The resulting front shape is 
$f(z) = -\alpha D \cos(z) + D^2 v_2 t+ D^2 f_2 \cos(2z)$ where $v_2$ and $f_2$ are rational functions of $a$ and $v$ \cite{Supplemental}. The point of maximum curvature is $z=0$ and there
$f''(z=0) = \alpha D -4 D^2 f_2$, so the front is focusing when $f_2<0$ and defocusing otherwise.
Analysis shows that $f_2$ becomes negative when the dimensionless parameter $\sqrt{1-v^2}\,\mathrm{d}\log\Gamma_0/\mathrm{d} v =a \sqrt{1-v^2} /(1+av) \gtrsim 1$. 
Thus, front curvature grows sub-linearly with $D$ when $\sqrt{1-v^2}\, \mathrm{d}\log\Gamma_0/\mathrm{d}v\lesssim 1$ and super-linearly otherwise.
An extensive study of the  $(a,v)$ parameter plane for the encounter of a front with a disk-shaped obstacle yielded the same trends \cite{Supplemental}. 

Can nonlinear self-focusing drive the transition from facet to micro-branch formation? As seen in Fig. \ref{fig1}(b) both structures are composed of step-lines that in one case diverge and in the other converge. To answer this question we use the results of our experimental studies of facet formation \cite{Kolvin.17}, which showed that the formation of a step incurs a local energetic cost of $\Gamma_0\delta A(z-z_0)$. Here $z_0$ is the position of the step along the front and $\delta A = (D/\pi)(1\pm\alpha ( z/w))/(1+( z/w)^2)$, where $\alpha=0.24$. 
During crack propagation steps drift and grow, leading to changes in position $z_0$ and width $w$. The width is assumed to be much smaller than system dimensions $w\ll Z$, so that $\delta A(z)$ is a sharply peaked distribution. Locally, a step-line forms a constant $\sim 45^\circ$ angle with the front slope \cite{Kolvin.17}. To numerically propagate the distribution $\delta A$ we draw a $45^\circ$ line from $z_0$ at a given time step and its intersection with the front at the next time step determines the new value of $z_0$.
 
\begin{figure}[h!]
	\includegraphics[scale=1]{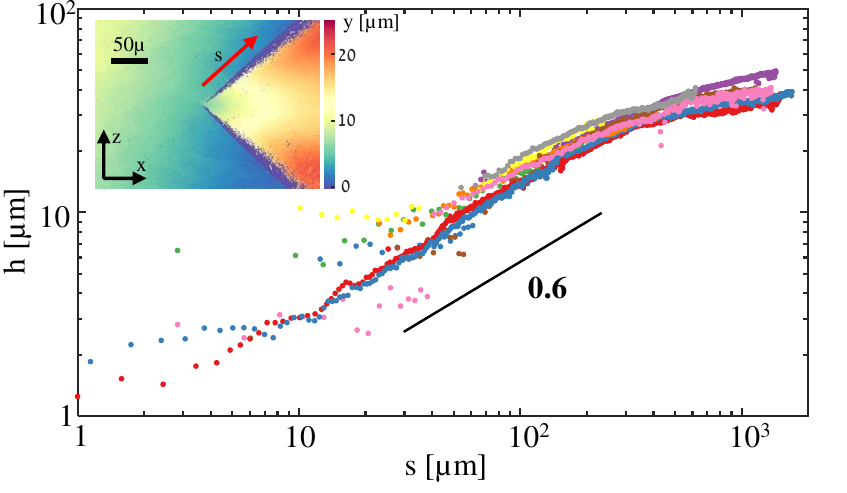}
	\caption{The growth of step height along the step-line backbone $s$ taken from pair nucleation events. Data are shown from 9 step-lines taken from 5 events. For two of the step-lines (red and blue points) we performed a high resolution measurement (at 0.5$\mu m/pix$). (inset) The high resolution surface scan.}\label{fig3}
\end{figure}

Lastly, we need to specify how steps grows with crack propagation. To this end, we consider the measured growth of step height $h$ obtained by profilometry on surfaces formed by the tensile fracture of polyacrylamide gel (for details, see \cite{Kolvin.17}).
While step growth is highly sensitive to the presence of neighboring step-lines 
and sample boundaries, we find that steps nucleating in pairs at a sufficient distance from other structures follow a reproducible trend, as shown in Fig. \ref{fig3}. Step heights grow along the step-line backbone $s$ as $h/\xi\sim(s/\xi)^b$ with $b=0.6\pm 0.1$ and $\xi=2\pm 1 \mu m$. Assuming that step widths and heights grow in proportion to each other, we take $w = (\xi(x+\xi))^{0.5}$ where $x$ is the position of the front at time $t$ and $z=z_0$; the distribution $\delta A$ has a width $w=\xi$ at $t=0$. 

We  numerically solve Eq. (\ref{energyreleaserate}) for two initially diverging step-lines with the parameters $v=0.1, a=4, \xi=0.0016Z$. Step-line centers were initially separated by $10\xi$ and the front was straight. 

\begin{figure}[h!]
	\includegraphics[scale=1]{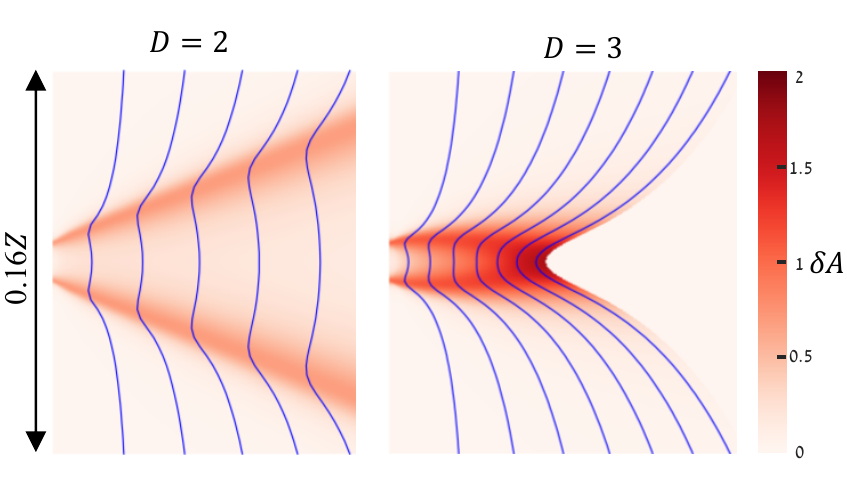}
	\caption{Transition from step-line divergence (\textit{left}) to step-line convergence (\textit{right}) when increasing the dissipation amplitude $D$. Here $v=0.1$, $a=4$ and $\xi = 0.0016Z$. This is the same qualitative behavior as observed experimentally in Fig. \ref{fig1}(b). See movies in the supplemental material.}\label{fig4}
\end{figure}

The left panel of Fig. \ref{fig4} depicts crack front and local fracture energy evolution for an amplitude of $D=2$. The two step-lines diverge with an angle that continually increases during propagation. Increasing the amplitude of $\delta A$ to $D=3$ causes the step-lines to converge instead of diverging (see the right panel of Fig. \ref{fig4}). In our previous study \cite{Kolvin.17} we have seen that $\int dz \delta A \sim 1.4 h$. For $w\sim h/2$ this relation translates into $D\sim 2.8$. This suggests that convergence in the simulation may occur within the same parameter range as observed experimentally. 

Additional simulations show similar trends; decreasing $a$ and increasing $v$ pushes the transition to step-line convergence to higher values of $D$. Changing $\xi$, however, does not affect the transition, but only the overall scale of the process. When solving Eq. (\ref{energyreleaserate}) under the same conditions as Fig. \ref{fig4}, but neglecting 2\textsuperscript{nd} order terms, no convergence was observed.


How general is the nonlinear focusing observed here?
The crack front equation of motion presented here was derived in the context of a scalar approximation to elasticity and based on the assumption of planarity. We believe, however, that our results are not constrained by these assumptions.
Perturbations of the crack front in both scalar and vector models generically decay. 
When $a=0$, the wave-like modes that transmit stress along the crack front decay as $1/\sqrt{t}$ in the scalar model following an encounter with an asperity; in Mode I elastodynamics wave amplitudes undergo a short decay phase followed by a fixed value (``front waves'') \cite{Morrissey.98}. 
However, for $a>0$ these modes decay exponentially in both theories. 
Moreover, the dimensionless parameter $\sqrt{1-v^2}\,\mathrm{d}\log\Gamma_0/\mathrm{d} v$ will arise in any nonlinear treatment of crack front perturbations, independent of the model used.
Planarity, on the other hand, appears to be a good approximation even when the crack front is composed of disconnected branches \cite{Kolvin.17}. The main effect of an out-of-plane excursion on in-plane dynamics is the local increase in dissipation due to the additional surface area formed by the crack. However, the full 3D dynamics must become relevant once two step-lines converge and meet. 

What are the implications of our results to the microbranching instability?
Micro-branching is a complex phenomenon, and some of its features are material dependent.
Our findings suggest that micro-branch localization in $z$ results from the development of high in-plane curvature along the crack front. Our model predicts how local curvature is controlled by three parameters: the local dissipation $\delta A$, the crack velocity $v$ and $\d{\log\Gamma_0}{v}=a/(1+av)$. In the linear limit $\delta A \ll 1$; $\delta A$ increases curvature while $v$ diminishes it. When $\delta A\sim 1$ nonlinear effects become important. For $\sqrt{1-v^2}\,\mathrm{d}\log\Gamma_0/\mathrm{d} v \leq 1$ curvature increase with dissipation is sub-linear while for $\sqrt{1-v^2}\,\mathrm{d}\log\Gamma_0/\mathrm{d} v \geq 1$ it is super-linear.
In experiments $\delta A$ and $v$ are not independent; as $v$ increases, surface and sub-surface structure becomes more complex leading to dynamically increased dissipation with $v$. This effective increase in $\delta A$ may be significant, possibly explaining why experiments \cite{Kolvin.17,Goldman.15} indicate that faster cracks are more susceptible to out-of-plane perturbations.

A previously unappreciated dimensionless parameter, $\sqrt{1-v^2}\,\mathrm{d}\log\Gamma_0/\mathrm{d} v$, that controls the sign and magnitude of nonlinear focusing, may explain at least part of the variation in micro-branching between materials. 
For example, self-focusing might be the reason why in polyacrylamide gels, where $a\sim 3.3$, micro-branches appear at $v\sim 0.1 c_R$, and in soda-lime glass, where $a\sim 0$, micro-branches appear at $v\sim 0.4 c_R$.
Further study of crack front dynamics in the presence of nonlinear elasticity \cite{Bouchbinder.08,Bouchbinder.14} as well of the nucleation and growth of surface structure is needed to clarify our understanding of the micro-branching transition.

\textit{Acknowledgements.}
J.F. and I.K. acknowledge the support of the Israel Science Foundation (grant no.1523/15), as well as the US-Israel Bi-national Science Foundation (grant no. 2016950). 
\bibliography{bibl}

\end{document}


\title{Supplemental Material for ``Nonlinear focusing in dynamic crack fronts and the micro-branching transition"}
\author{Itamar Kolvin$^1$, Jay Fineberg$^1$ and Mokhtar Adda-Bedia$^2$}

\affiliation{$^1$ Racah Institute of Physics, The Hebrew University of Jerusalem, Jerusalem, Israel 9190401 \\
$^2$ Universit\'{e} Lyon, Ecole Normale Sup\'{e}rieure de Lyon, Universit\'{e} Claude Bernard, CNRS, Laboratoire de Physique, F-69342 Lyon, France}
\maketitle
\section{Explicit 2\textsuperscript{nd} order equation of motion for scalar crack fronts}
Here we derive the explicit equation of motion used to propagate the crack front in our simulations. In the following, we define the scalar wave speed $c=1$ and $\alpha = \sqrt{1-v^2}$. Combining Eq. (1) and (3) of the main text, local energy balance reads
\begin{equation}\label{energy_balance}
G_0 g(v_\perp)(1+H[f]) = \tilde{\Gamma}_0(1+av_\perp)(1+\delta A)\,,
\end{equation}
where the instantaneous position of the crack front is given by $x = vt+f(z,t)$.
Approximating  $v_\perp\simeq v+f_t-\frac{v}{2}f_z^2$, where subscripts denote derivatives,  and keeping only terms up to the 2\textsuperscript{nd} order in $f$ and $\delta A$, Eq. (\ref{energy_balance}) becomes
\begin{equation}\label{energy_balance_expanded}
\left(1-\frac{1}{\alpha^2}\left(f_t-\frac{v}{2}f_z^2\right)+\frac{1-2v}{2\alpha^4}f_t^2\right)\left(1+H[f]\right) = \left(1+\frac{a}{1+av}\left(f_t-\frac{v}{2}f_z^2\right)\right)(1+\delta A)\,,
\end{equation}
where we assumed that for the unperturbed crack, $G_0 g(v) =\tilde{\Gamma}_0(1+av)$.
To obtain an explicit expression for $f_t$, we substitute $H[f]$ with the right-hand-side of Eq. (2) and expand the parentheses in (\ref{energy_balance_expanded}). The resulting equation is quadratic in $f_t$, which we solve, retaining only terms that are up to 2\textsuperscript{nd} order in $f$. The result is an explicit equation of motion for $f$:
\begin{equation}\label{explicit_eom}
\begin{split}
f_t =& \frac{v}{2}f_z^2+\frac{1}{1+\chi}\left(-\Psi[f]-\frac{1+2\chi-\chi^2+4v}{4\alpha^2(1+\chi)^2}\Psi[f]^2+\frac{1}{2\alpha^2}\Psi[f\Psi[f]]-\frac{1-2v}{4\alpha^2}\Psi_2[f^2]-\frac{1+2v}{2\alpha^2}f\Psi_2[f]\right)\\
&+\frac{1}{1+\chi}\left(-\alpha^2\delta A+\frac{\chi^2-2v}{(1+\chi)^2}\Psi[f]\delta A+\frac{2\chi(1-\chi)+1-2v}{2(1+\chi)^2}\alpha^2\delta A^2\right)\,.
\end{split}
\end{equation}
$\Psi_2[f]$ is defined before Eq. (2) in the main text.

\section{Details of the numerical solution}

We discretize Eq. (\ref{explicit_eom}) in both $z$ and $t$, assuming periodic boundary conditions. Since the problem does not contain an intrinsic length-scale we conveniently set the periodic $z$ interval length to $Z=2\pi$. 
The crack front is propagated in a straightforward Euler scheme. 
For a $z$-mesh of $N$ points, the time step is defined as $\Delta = \frac{0.2}{N\alpha}$. This time step is sufficiently small to finely sample the oscillations of the Bessel functions appearing in the history functionals $\Psi$ and $\Psi_2$ for the highest wavenumber $k_{max} = N/2$.
The history functionals are evaluated in $k$-space by trapezoidal integration with the same time step $\Delta$ used to propagate the crack front, and then transformed back to $z$-space with a Fast Fourier Transform. This integration scheme results in an error of $\sim 0.1\%$.
To prevent negative crack front velocities, we equated $f_t=-v$ locally when Eq. (\ref{explicit_eom}) gives $v+f_t<0$.

\textit{Disk-shaped obstacle --- Fig. 1(d) and Fig. 2.}
For the crack front interaction with an obstacle, we assume that the front is straight and initially positioned at $x=0$. The $z$ interval is discretized over $N=512$ mesh points.
We insert an disk-shaped obstacle of diameter $d = 0.05\pi$ centered at $x = d/2+2\epsilon$ where $\epsilon = 20\pi/N$. 
Inside the obstacle diameter $\delta A = D=const.$ and outside $\delta A = D\exp(-2(r-d/2)^2/\epsilon^2)$, where $r$ is the radial distance from the center of the obstacle.

\textit{Step-lines --- Fig. 4.}
To speed up simulations while maintaining numerical accuracy, the crack front was discretized on a $N=4096$ mesh which was made coarser at predetermined times as $\delta A$ grew wider; we doubled the mesh size at $t_N = (10\pi/2^N)^2/\xi v - \xi/v$ which roughly corresponds to the times when $w$ increased past 5 times the mesh size. 
Both crack front velocity and curvature evolved smoothly during the simulation, ensuring that re-meshing did not affect the overall dynamics. 

\section{Focusing for $\delta A = D\cos(z)$ and for the disk obstacle}

Eq. (\ref{explicit_eom}) may be solved analytically for the time-independent $\delta A = D\cos(z)$. 
Assuming that the spatial variation of the crack front is also time-independent, $\Psi[f]\rightarrow \alpha \mathcal{H}[f_z],\,\Psi_2[f]\rightarrow-\alpha^2 f_{zz}/2$, defining the Hilbert transform $\mathcal{H}[g] = \pi^{-1}\int \frac{dz'}{z-z'}g(z')$. Then, Eq. (\ref{explicit_eom}) reduces to
 \begin{equation}\label{explicit_time_independent_eom}
 \begin{split}
 (1+\chi)f_t + \alpha \mathcal{H}[f_z] +\alpha^2\delta A = &-\frac{1+2\chi-\chi^2+4v}{4(1+\chi)^2}\mathcal{H}[f_z]^2+\frac{1}{2}\mathcal{H}[\partial_z(f\mathcal{H}[f_z])]+\frac{1+2\chi v}{4}f_z^2+\frac{1}{4}f f_{zz}\\
&+\frac{\chi^2-2v}{(1+\chi)^2}\alpha \mathcal{H}[f_z]\delta A+\frac{2\chi(1-\chi)+1-2v}{2(1+\chi)^2}\alpha^2\delta A^2\,.
 \end{split}
 \end{equation}
Substituting $f = D f_1 \cos(z) + D^2 v_2 t + D^2f_2\cos(2z)$ in Eq. (\ref{explicit_time_independent_eom}) and solving for order-by-order we obtain the coefficients
\begin{equation}
f_1= -\alpha;\; v_2 =  \alpha^2\frac{\chi  \left(-4 \chi +v (1+\chi )^2\right)}{4 (1+\chi )^3};\; f_2 = \alpha\frac{1+(2-v) \chi -(3+2 v) \chi ^2-v \chi ^3}{8 (1+\chi )^2}\,.
\end{equation}

\begin{figure}[h!]
	\includegraphics[scale=1]{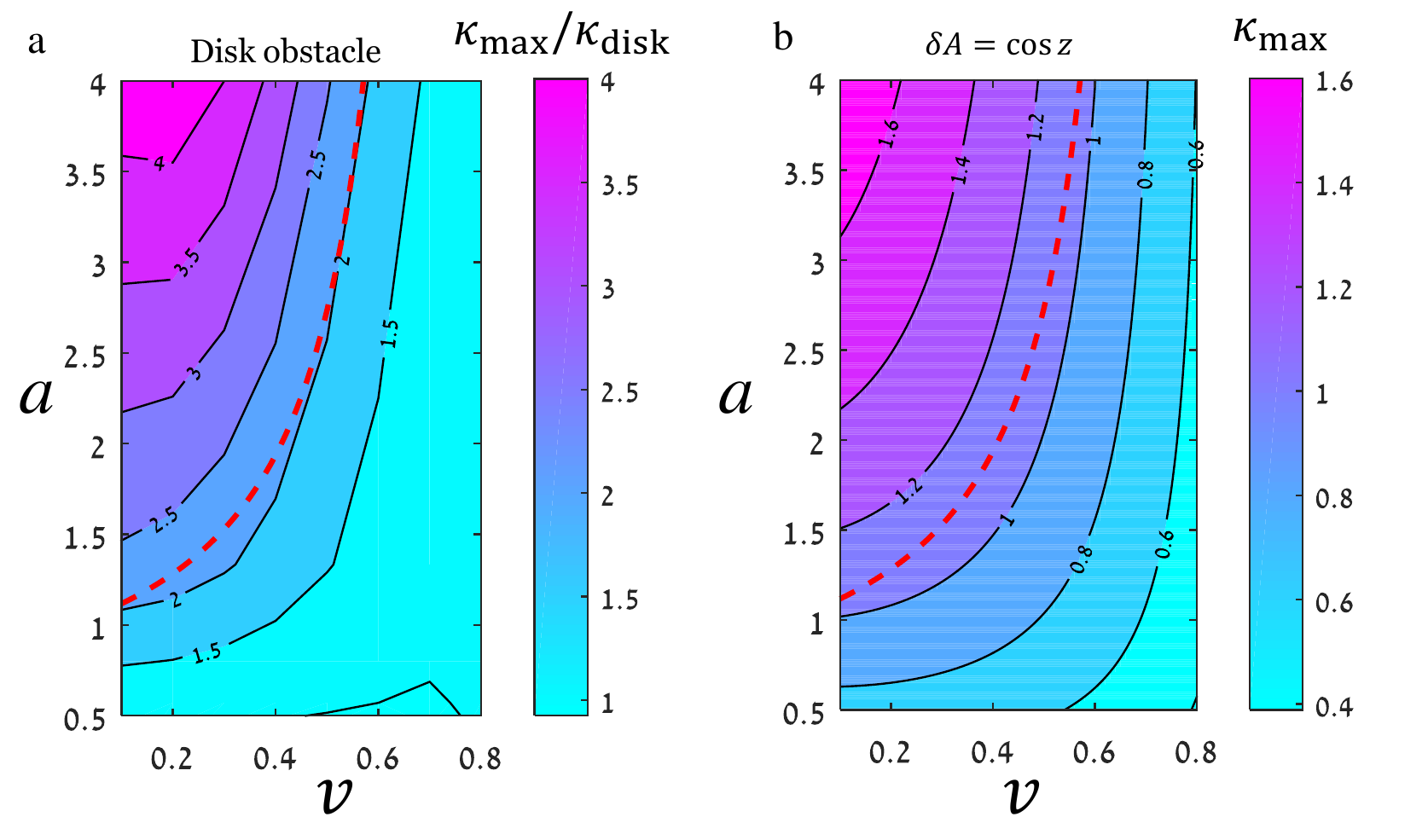}
	\caption{Comparison of maximum curvatures in (a) interaction of a crack front with the disk obstacle and (b) the time independent $\delta A = D\cos(z)$. In both cases $D=1$. The disk curvature is $\kappa_{disk} = 2/(0.05\pi)$. The red dashed line corresponds to $\sqrt{1-v^2}a/(1+av)=1$ and indicates the transition from defocusing to focusing.}\label{fig_S1}
\end{figure}

In Fig. \ref{fig_S1} we compare the maximum curvature obtained by this solution with that observed during the dynamic interaction with the disk obstacle. The curvatures in the two cases show similar dependencies on $a$ and $v$, as well as a transition from nonlinear defocusing to focusing when $\sqrt{1-v^2}a/(1+av)\sim 1$.

\section{Supplemental movie files}
\textit{fig1b\_movie.avi} --- The movie shows the propagation of crack fronts depicted in Fig. 1(b) and contains 190 frames originally taken at 15000 fps (total duration: 12.7ms), displayed here at 20fps.

\textit{fig4left\_movie.avi, fig4right\_movie.avi} --- The two movie files show the crack front dynamics depicted in Fig. 4 of the main text. The duration of the first movie is $T=14.09$ and the duration of the second movie is $T=14.82$, where we take $Z=2\pi$ and $c=1$.